\documentclass[aps,twocolumn,floatfix,showpacs]{revtex4}
\usepackage{graphicx,bm}
\usepackage{amsmath}
\usepackage{amssymb}
\usepackage{mathbbol}

\begin{document}
\title{Crossover in the Efimov spectrum}
\author{Ludovic Pricoupenko}
\affiliation
{Laboratoire de Physique Th\'{e}orique de la Mati\`{e}re Condens\'{e}e, 
Universit\'{e} Pierre et Marie Curie and CNRS,
4 place Jussieu, 75252 Paris Cedex 05, France.}
\date{\today}
\begin{abstract}
A filtering method is introduced for solving the zero-range three-boson  problem. This scheme permits to solve the original Skorniakov Ter-Martirosian integral equation for an arbitrary large ultraviolet cut-off and avoiding the Thomas collapse of the three particles. The method is applied to a more general zero-range model including a finite background two-body scattering length and the effective range. A cross-over in the Efimov spectrum is found in such systems and a specific regime emerges where Efimov states are long-lived.
\end{abstract}
%
\maketitle

\section*{INTRODUCTION}

In bosonic systems where two-body scattering is resonant, Efimov predicted in the early 70's the existence of shallow trimers with a spectrum characterized by a discrete scaling symmetry and an accumulation point at zero energy \cite{Efi70}. The first clear evidence of such states occurred in ultra-cold atomic systems \cite{Kra06,Kno09,Zac09,Noa09}, where  fine control of the resonant behavior through a Feshbach resonance (FR) mechanism is possible \cite{Chi10}. The few-body problem which was originally deeply linked to nuclear physics is now a hot topic in ultra-cold physics \cite{Ste09,Fer09,Pol09} and plays a significant role also in the search for  highly correlated many-body states \cite{Lev07,Pet07,Lev09}. While experiments unambiguously confirm general properties of Efimov physics \cite{Bra06}, deviations from universal predictions are observed supporting the interest in using detailed approaches like the one in Ref.~\cite{Din09} or including a microscopic description of the FR mechanism with a coupling between atoms (in the 'open channel') and molecules (in the 'closed channel') as in Refs.~\cite{Lee07,Mas08,Jon10}. In the latter references, the finite width of the magnetic FR and the 'background' scattering length (denoted ${\Delta \mathcal B}$ and ${a_{\rm bg}}$, respectively) appear as key parameters for the low energy scattering amplitude which plays a central role in three-body properties \cite{Mas08,Jon10}. At the two-body level, two-channel models permit the expression of the scattering length (denoted $a$) as a function of the external magnetic field ${\mathcal B}$ in the vicinity of a given resonance located at ${\mathcal B=\mathcal B_0}$ with:
\begin{equation}
a = a_{\rm bg} \left( 1 - \frac{\Delta \mathcal B}{\mathcal B - \mathcal B_0} \right).
\end{equation}
The magnetic width can also be characterized by a length ${R^\star}$ hereafter referred as the 'width radius' and defined by ${R^\star = \hbar^2/(m a_{\rm bg} \delta \mu \Delta \mathcal B)}$, where $\delta \mu$ is the difference in magnetic moment for an atomic pair in the open- and in the closed-channel \cite{Pet04b}. In the limit of an asymptotically narrow resonance, the width radius takes a large value as compared to the range of the interparticles forces (denoted by ${b}$) and defines a low energy scale. In this peculiar regime it is possible to determine Efimov physics and deviations from universality (in the intermediate detuning regime) both in terms of $a$ and $R^\star$ only \cite{Pet04b,Gog08}. Recently, it has been shown that the superposition of an external electric field permits to tune the background scattering length ${a_{\rm bg}}$  and thus the width radius ${R^\star}$ \cite{Mar08} which makes possible study of low energy three-body properties as a function of these quantities.

Following this idea, a zero-range model generalizing the universal theory of broad resonances and the effective range approach of ultra-narrow resonances is solved in what follows. This modeling is especially pertinent in situations where each of the three parameters $(a,a_{\rm bg},R^\star)$ of the FR are large (in absolute value) with respect to the range $b$ which is of the order of the van der Waals radius given by ${\left(mC_6/\hbar^2\right)^{1/4}}$ (where $C_6$ is the van der Waals coefficient of the interatomic potential). In this model, the {\it so-called} three-body parameter (denoted ${\kappa^\star}$) which characterizes the high momentum of the three-body wave function, is introduced by means of a nodal condition. While keeping the simplicity of the zero-range approach, this 'filtering' method of physical solutions permits avoiding the Thomas collapse \cite{Tho35,Min62}
. In the limit of broad resonances, the nodal condition implemented in the original Skorniakov Ter-Martirosian (STM) equation (Eq.(12) in Ref.~\cite{Sko57}) through a simple subtraction scheme permits recovery of the results of the 'universal theory'. In the limit of asymptotically narrow resonances ${(R^\star\to \infty)}$, a decoupling  occurs between shallow states of the effective range theory \cite{Pet04b,Gog08} and deeper states depending on ${\kappa^\star}$. Between the ultra-narrow  and the infinitely broad resonance limits, the Efimov spectrum exhibits a crossover involving these two families of states. In the regime of large positive background scattering length and width radius, the model supports a shallow dimer state. In this situation, Efimov trimers of lowest energy are quasi-bound states and manifest themselves as asymmetric resonance peaks in atom-dimer scattering which are characteristic of Fano resonances \cite{Fan61}. This result confirms the scenario introduced in Nuclear physics for the compound system made of two neutrons interacting with one {${^{18}}$C} nucleus \cite{Maz06}.

\section{Model for two interacting particles}

The zero-range approach used hereafter is based on the expression of the low energy two-body $s$-wave scattering phase shift (denoted ${\delta(k)}$) or equivalently of the scattering amplitude (denoted by ${f(k)}$), obtained from the limit of zero potential range (${b\to0}$) of 
realistic models \cite{Mar08,Jon10,Wer09a} or directly from a zero-range diagrammatic approach (see Eq.~(2) in Ref.~\cite{Mas08}). In this limit, the scattering phase shift (related to the scattering amplitude by ${\frac{1}{f(k)}=k \cot \delta(k)-ik}$) at collisional momentum $k$ is: 
\begin{equation}
\cot \delta(k) =-\frac{1}{a k}-\frac{R^\star(1-\frac{a_{\rm bg}}{a})^2k}{R^\star a_{\rm bg}(1-\frac{a_{\rm bg}}{a})k^2+1} . 
\label{eq:phase_shift}
\end{equation}
Equation~\eqref{eq:phase_shift} can also be obtained in the low momentum limit (${k b\ll 1}$) of the models in Refs.~\cite{Mar08,Jon10}, and is thus the basic object which describes the interparticle interaction for low energy process \cite{LowE}. For a vanishing background scattering length, Eq.~\eqref{eq:phase_shift} coincides with the effective range approximation used in Refs.~\cite{Pet04b,Gog08}.  Systems studied hereafter are translation invariant, hence the zero-range approach is expressed directly in the momentum representation in the same manner as in Refs.~\cite{Pri08b,Pri10a}. In the center of mass frame, the wave function of the two interacting particles of mass $m$ at colliding energy ${E_{\rm col}=\hbar^2k_{\rm col}^2/m}$ satisfies
\begin{equation}
\left(\frac{\hbar^2}{m} k^2 - E_{\rm col}\right) \langle \mathbf k |\Psi \rangle = 
\frac{4\pi \hbar^2 S_{\Psi}}{m} ,
\label{eq:2body}
\end{equation}
where $k$ is the relative momentum. The zero-range interaction appears as a source term in the right hand side of Eq.~\eqref{eq:2body} and the source amplitude (denoted ${\rm S_\Psi}$) is obtained from a contact condition in the configuration space which reduces to an integral equation in the momentum space
\begin{equation}
{\rm Reg}_{\epsilon \to 0} 
\int \frac{d^3 {\mathbf k}}{(2\pi)^3} e^{-k^2 \epsilon^2} \langle {\mathbf k} | \Psi \rangle 
= \frac{k_{\rm col}{\rm S}_\Psi}{\tan \delta(k_{\rm col})},
\label{eq:contact}
\end{equation}
where the operator ${{\rm Reg}_{\epsilon \to 0}}$ extracts the regular part of the integral in the limit $\epsilon \to 0$. In the regime where $R^\star$ is vanishingly small, the condition expressed in Eq.~\eqref{eq:contact} is equivalent to the Bethe-Peierls contact condition introduced in pioneering works on the deuteron \cite{Bet35}.

\section{Three interacting bosons}
\subsection{Filtering the generalized Skorniakov Ter-Martirosian equation}

The formalism is now applied to three identical bosons of mass $m$ at negative energy ${(E)}$ in the center of mass frame. The wave function of the three particles labeled ${i}$ (${i\in \{ 1,2,3\}}$) with respective momentum ${\mathbf k_i}$ factorizes as $(2\pi)^3\delta(\sum_i \mathbf k_i) \psi( \mathbf k_1,\mathbf k_2,\mathbf k_3)$ and
\begin{equation}
\left( \sum_{i=1}^3 \frac{k_i^2}{2} + q^2 \right)    \psi( \mathbf k_1,\mathbf k_2,\mathbf k_3) = 4\pi \sum_{i=1}^3 \phi(\mathbf k_i),
\end{equation}
where $q$ is a positive wavenumber defined by ${E=-\hbar^2q^2/m}$ and ${\phi(\mathbf k_i)}$ is the source amplitude associated with the pair ${(jk)}$ (${i,j,k}$ are distinct labels). The contact condition in Eq.~\eqref{eq:contact} applied for the pair ${(23)}$ with the colliding energy ${E_{\rm col}=E-\frac{3\hbar^2}{4m} \mathbf k_1^2<0}$ gives the integral equation satisfied by the source amplitude:
\begin{equation}
\frac{\phi({\bf k})}{f(iq_{\rm col})} = 8\pi \int \frac{d^3{\bf u}}{(2\pi)^3}\, \frac{ \phi({\bf u})}{u^2 +k^2 + {\bf k}.{\bf u} +q^2},
\label{eq:Skorniakov}
\end{equation}
where ${q_{\rm col} = \sqrt{q^2+\frac{3k^2}{4}}}$ is the imaginary colliding momentum. Equation~\eqref{eq:Skorniakov} is a generalization of the STM equation (corresponding to the case ${R^\star=a_{\rm bg}=0}$). It was first derived by Massignan and Stoof (Eqs.~(1) and (2) in Ref.~\cite{Mas08}) and follows from the zero-range limit of the generalized STM equation in Ref.~\cite{Jon10}. From these two references, ${\phi(k)}$ can be interpreted as a dressed (atom $\otimes$ closed channel molecule) wave function (the molecule in the closed channel is not described by the zero range approach). For ${R^\star\ne 0}$ and ${a_{\rm bg}=0}$ (\emph{i.e.} the FR is asymptotically narrow), Eq.~\eqref{eq:Skorniakov} permits extraction of all the low energy three-body properties and it has been solved in Refs.~\cite{Pet04b,Gog08}. For  ${a_{\rm bg} \ne 0}$ or ${R^\star= 0}$, Eq.~\eqref{eq:Skorniakov} does not constitute a well defined problem in the $s$-wave sector of the source amplitude. This can be shown along the same lines as in Ref.~\cite{Dan61} by considering the high momentum behavior of the eigenfunctions in the $s$-wave sector. In this limit, the scattering amplitude coincides with the unitary expression, ${f(iq_{\rm col}) \to 1/q_{\rm col}}$ and Eq.~\eqref{eq:Skorniakov} supports a pair of power-law solutions of the form ${\phi(k) \sim k^{-2\pm is_0}}$ where ${s_0 \simeq 1.00624}$. Danilov showed that the STM model is self-adjoint if one filters eigenfunctions by fixing the linear combination of the two conjugate solutions as ${k\to \infty}$ with an asymptotic phase shift. However the spectrum obtained from this filtering procedure is not bounded from below \cite{Min62}: this is the Thomas collapse \cite{Tho35} analog to the falling of a particle in a $1/r^2$ attractive potential \cite{Efi70}. In the present work, instead of the Danilov's phase-shift filtering, a nodal condition is used which permits getting a finite minimal energy in the spectrum. This method is based on the expression of the source amplitude at unitarity (\emph{i.e} ${R^\star=0}$ and ${|a|=\infty}$), which is given, up to a normalizing constant by \cite{Gog08}:
\begin{equation}
\phi(k) = \frac{1}{k \sqrt{q^2+3k^2/4}} \sin \left[ s_0 \operatornamewithlimits{arcsinh}\left(\frac{k\sqrt{3}}{2q}\right)\right] .
\label{eq:exact-unitary}
\end{equation}
Equation~\eqref{eq:exact-unitary} supports the discrete scaling symmetry found by Efimov and the binding wavenumbers verify:
\begin{equation} 
q_n=\kappa^\star e^{-n\pi/s_0} ,
\label{eq:bindingwavenumber} 
\end{equation}
where ${n}$ is an integer and ${\kappa^\star}$ is the three-body parameter. At a large momentum, one obtains: 
\begin{equation}
\phi(k) \operatornamewithlimits{\sim}_{k \gg q} \frac{1}{k^2} \sin \left[ s_0 \ln\left(\frac{k\sqrt{3}}{\kappa^\star} \right)\right] ,
\label{eq:def_kappa*}
\end{equation}
which gives the relation between the three-body parameter and the Danilov's phase shift.  For $R^\star=0$, the spectrum of the zero-range (or universal) theory is invariant in the transformation ${\kappa^\star \to \kappa^\star\exp(\pi/s_0)}$ showing that $\kappa^\star$ is not unique. However, for interatomic forces of finite range $b$, the binding wavenumber $q_n$ in Eq.~\eqref{eq:bindingwavenumber} is bounded from below by $1/b$. In what follows, $\kappa^\star$  is chosen to be of the order of ${1/b}$ meaning that the scaling law in Eq.~\eqref{eq:bindingwavenumber} is supposed to be valid for ${n \ge 0}$. Unphysical solutions of Eq.~\eqref{eq:Skorniakov} are characterized by a large binding wave number ${(qb \gtrsim 1)}$ and are also given by Eq.~\eqref{eq:exact-unitary} for the zero-range theory at unitarity. By noting that as $k$ increase from $0$, the first node in the expression of Eq.~\eqref{eq:exact-unitary} is found at a momentum of the order of $q$, it is possible to filter out the unphysical eigenfunctions of the STM equation by imposing the nodal condition:
\begin{equation}
\phi(k_{\rm reg})= 0 \quad \mbox{with} \quad  k_{\rm reg}= \frac{\kappa^\star}{\sqrt{3}} e^{\pi/s_0} . 
\label{eq:filtre} 
\end{equation}
That way the spectrum obtained from the STM equation at unitarity has a minimum energy ${E_0 \simeq -\hbar^2\kappa^\star\,^2/m}$ (with a relative error given by ${2\exp(-2\pi/s_0)\simeq 3.8 \times 10^{-3}}$) and for ${E_n>E_0}$, it coincides asymptotically (\emph{i.e.} for large $n$) with the Efimov spectrum of the 'universal theory' \cite{Efi70,Bra06}. 

In the regime where ${1/a\ne 0}$ and/or ${(a_{\rm bg} \ne 0, R^\star \ne 0)}$, the three-body parameter $\kappa^\star$ characterizes the high-momentum behavior of the eigenstates (\emph{i.e.} the Danilov's phase-shift) given by Eq.~\eqref{eq:def_kappa*} which is a supplementary and necessary contact condition at the three-body level. In this regime of parameters, at least if a trimer exists, the binding wave numbers no longer verify the scaling law Eq.~\eqref{eq:bindingwavenumber}. Formally, the fact that $\kappa^\star$ is a parameter which is independent of the low energy two-body properties (and thus of the width radius $R^\star$) can be seen as resulting from the high momentum behavior of the solution of the generalized STM equation (for finite $a_{\rm bg}$) which is of the same type as the universal solutions. In the specific limit where ${R^\star \ne 0}$ and ${a_{\rm bg} \to 0}$ in Eq.~\eqref{eq:phase_shift}, eigenfunctions are non zero in the region of momentum less than  the order of $1/R^\star$. In this limit, results are thus independent of $\kappa^\star$ and the spectrum is characterized by another momentum hereafter denoted  ${\kappa^\star\,'=2.65/R^\star}$ \cite{Pet04b,Gog08} and also called 'three-body parameter' in the latter references. In the general case where the parameters in Eq.~\eqref{eq:phase_shift} are finite, the nodal condition in Eq.~\eqref{eq:filtre} can again be used to prevent any Thomas collapse without modifying Eq.~\eqref{eq:Skorniakov}.
\begin{figure}[hx] 
\begin{center}
\resizebox{8 cm}{!}
{
\includegraphics{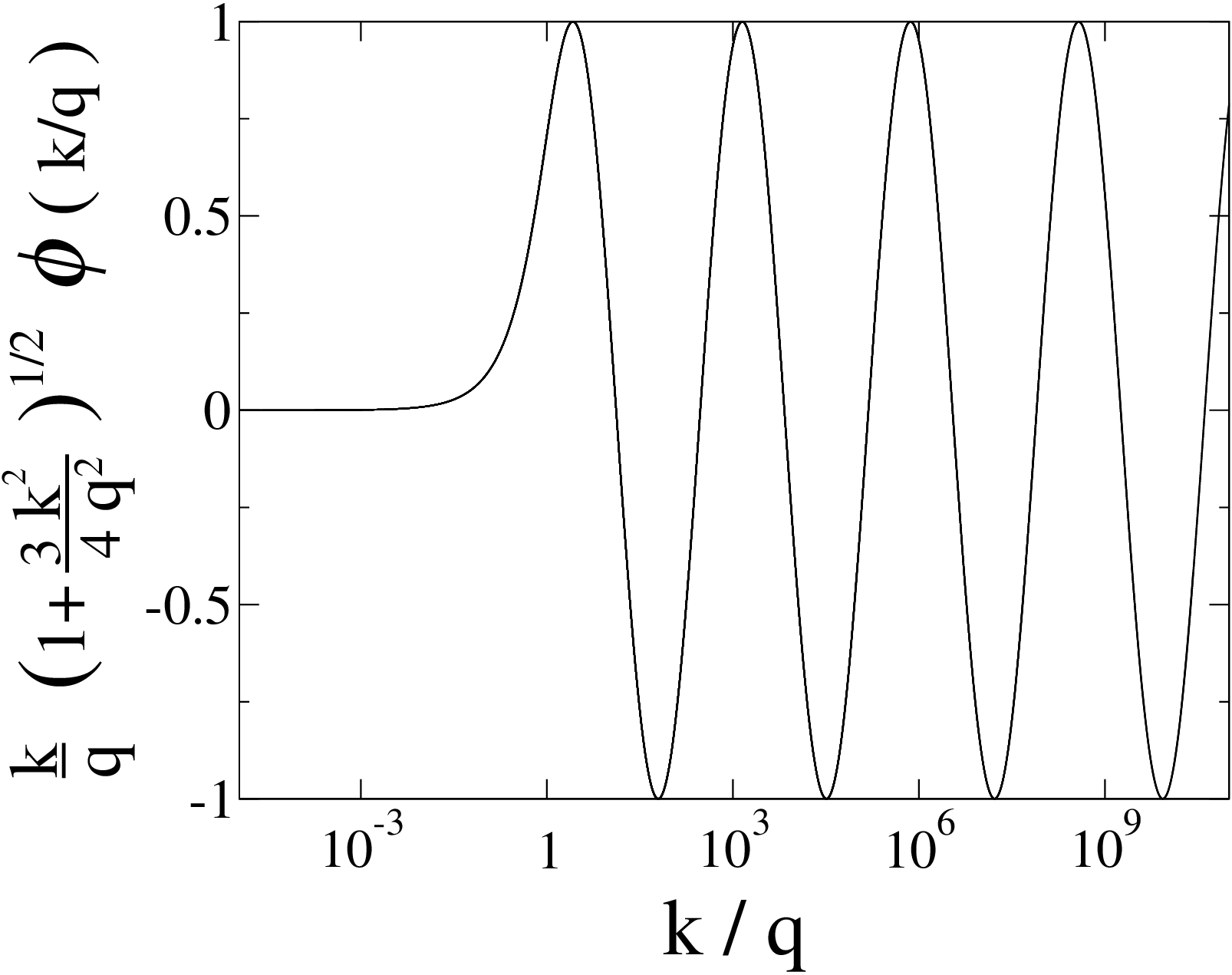}
}
\caption{Functions ${k/q\sqrt{1+3k^2/(4q^2)}\phi_n(k/q)}$ obtained at unitarity (${|a|=\infty, R^\star=0}$) from Eq.~\eqref{eq:sko_regular}  for the eight first trimers (including the ground state). The computation has been performed with the UV cut-off ${\Lambda_3=5\times10^2 \times \kappa^\star}$. The function 
${\sin \left\{ s_0 \operatornamewithlimits{arcsinh}[k\sqrt{3}/(2q)]\right\}}$ appearing in the exact solution Eq.~\eqref{eq:exact-unitary} has been superimposed and is not distinguishable from the functions obtained by solving Eq.~\eqref{eq:sko_regular}.}
\label{fig:function_univ}
\end{center}
\end{figure}

In numerical computations there always exists a large but finite high momentum cut-off (denoted hereafter ${\Lambda_3}$) in the evaluation of the right hand side of Eq.~\eqref{eq:Skorniakov} and one needs a stable procedure to implement the nodal condition Eq.~\eqref{eq:filtre}. This can be  achieved by subtracting term by term Eq.~\eqref{eq:Skorniakov} from its own expression at ${k=k_{\rm reg}}$ where Eq.~\eqref{eq:filtre} is also used. One obtains the regularized integral equation: 
\begin{multline}
\frac{\phi(k)}{f(iq_{\rm col})} = \frac{2}{\pi} \int_0^{\infty} \!\! du \,
\left[{\mathcal K}_q(k,u)-{\mathcal K}_q(k_{\rm reg},u) \right] \phi(u),
\label{eq:sko_regular}
\end{multline}
where the Kernel is given by:
\begin{equation}
{\mathcal K}_q(k,u)=\frac{u}{k} \ln \left(\frac{u^2+k^2+q^2+ku}{u^2+k^2+q^2-ku}\right).
\end{equation}
One can verify in numerical solutions of Eq.~\eqref{eq:sko_regular}, that results are invariant in a change of the UV cut-off ${\Lambda_3}$ of the integral term for ${\Lambda_3 \gg \kappa^\star}$. As a benchmark, the source amplitude ${\phi_n(k)}$ has been computed at unitarity for the eight first trimers with the UV cut-off $\Lambda_3=5\times10^2\times \kappa^\star$ and are displayed in Fig.~(\ref{fig:function_univ}). This figure illustrates the high degree of convergence of the method even for the first excited states (which is a consequence of the relatively large value of the scale factor ${e^{\pi/s_0}\sim 22.7}$). Moreover, the binding wavenumbers of the excited trimers differ from the exact values of Eq.~\eqref{eq:bindingwavenumber} by a relative error of less than $10^{-5}$.

While not equivalent, the present regularization procedure of the STM equation as some similarity to that used in the effective-field-theory approach to the three-boson problem \cite{Bed99} where an integral counter-term is also introduced to cancel the ${\Lambda_3}$ dependence. The present nodal condition has the advantage of making clear that the integral counter-term is just a way to achieve a filtering of the ``physical'' solutions which are eigenfunctions of the generalized STM equation Eq.~\eqref{eq:Skorniakov}. In Refs.~\cite{Ham01,Afn04} a subtraction scheme has been also introduced for the atom-dimer scattering equation [see Eq.~\eqref{eq:diff}] in the framework of the effective field theory. In the latter references, the renormalization is done at zero energy with the atom-dimer scattering length as the input parameter and this regularizing method does not refers to a nodal condition. Interestingly, this subtraction method was used in Ref.~\cite{Pla09} to evaluate linear effective range corrections to the universal predictions \cite{Bra06} near resonance ${(|a|\gg b)}$ in the regime where the effective range and $b$ are of the same order. In Ref.~\cite{Mas08} another regularizing technique was used: a supplementary effective range term ${-R_{\rm fit}^\star k}$ was introduced in the two-body phase shift of Eq.~\eqref{eq:phase_shift} with a fitting  parameter ${R_{\rm fit}^\star}$ of the order of $R^\star$, thus modifying the two-body low energy properties in the regime of a large width radius \cite{Rstar}. 

\subsection{Crossover in the trimer's spectrum}

\begin{figure}[b] 
\begin{center}
\resizebox{8 cm}{!}
{
\includegraphics{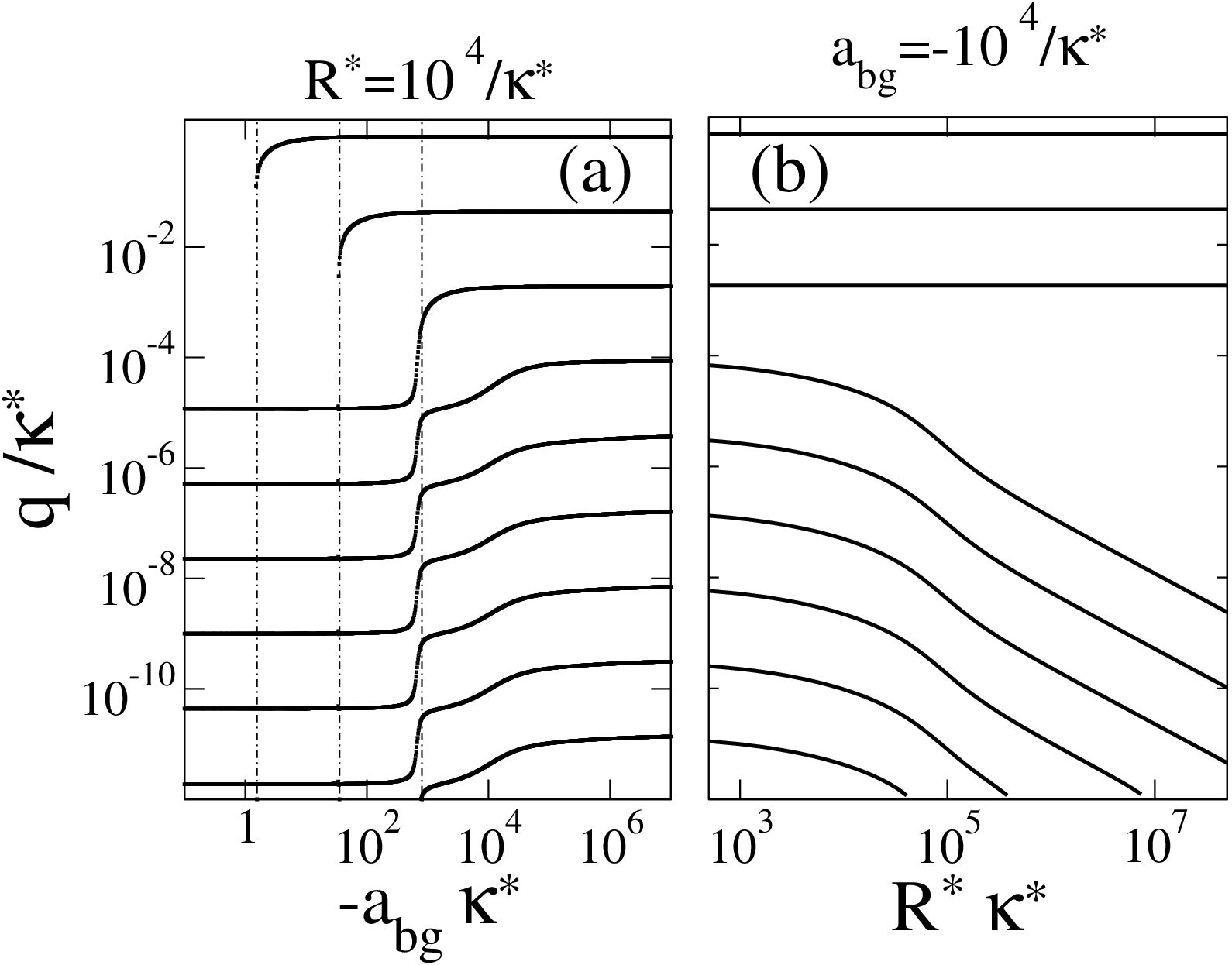}
}
\caption{Trimers' spectrum of Eq.~\eqref{eq:sko_regular} at resonance (${|a|=\infty}$) and for ${a_{\rm bg}<0}$. 
(a): Binding wavenumbers $q$ (continuous lines) as a function of ${a_{\rm bg}}$ for a constant and large value of the width radius $R^\star$; vertical dashed lines correspond to the thresholds ${a_{\rm bg} \kappa^\star \simeq-1.51 e^{{n\pi}/{s_0}}}$, ${n=0,1,2}$. This figure illustrates the crossover from a narrow FR (${R^\star \kappa^\star \gg 1}$ and ${|a_{\rm bg}| \ll R^\star}$) to a shape resonance (${|a_{\rm bg}|\kappa^\star}$ is large).
(b): Binding wavenumbers as a function of the width radius for a constant and large value of ${|a_{\rm bg}|\kappa^\star}$. This figure illustrates the crossover from a broad FR (${R^\star \kappa^\star \ll 1}$) to a narrow FR  superimposed on a shape resonance.}
\label{fig:Spec_Rstar_mabg}
\end{center}
\end{figure}
The present model can be used to exemplify the crossover in the Efimov spectrum as one varies the two parameters ${(a_{\rm  bg},R^\star)}$ going from an infinitely broad FR ${(R^\star=0)}$ to an asymptotically narrow FR ${(R^\star=\infty)}$. To focus the discussion on this issue, in what follows the model is solved at resonance  ${(|a|=\infty)}$. For a negative background scattering length ${(a_{\rm bg}<0)}$, there is no dimer, and the results are displayed in Fig.~(\ref{fig:Spec_Rstar_mabg}). This figure shows the crossover between two types of Efimov spectra, each supporting the discrete symmetry of scaling factor ${e^{\pi/s_0}}$ but having different ground states. The first one is the spectrum of the effective range approach~\cite{Pet04b} characterized by a ground state binding wave number ${q_0=\kappa^\star\,' e^{-\pi/s_0}}$ where ${\kappa^\star\,'\simeq 2.65/R^\star}$ ~\cite{Pet04b,Gog08}. The second is the spectrum of a broad resonance where ${q_0=\kappa^\star}$. 
\begin{figure}[hx] 
\begin{center}
\resizebox{8 cm}{!}
{
\includegraphics{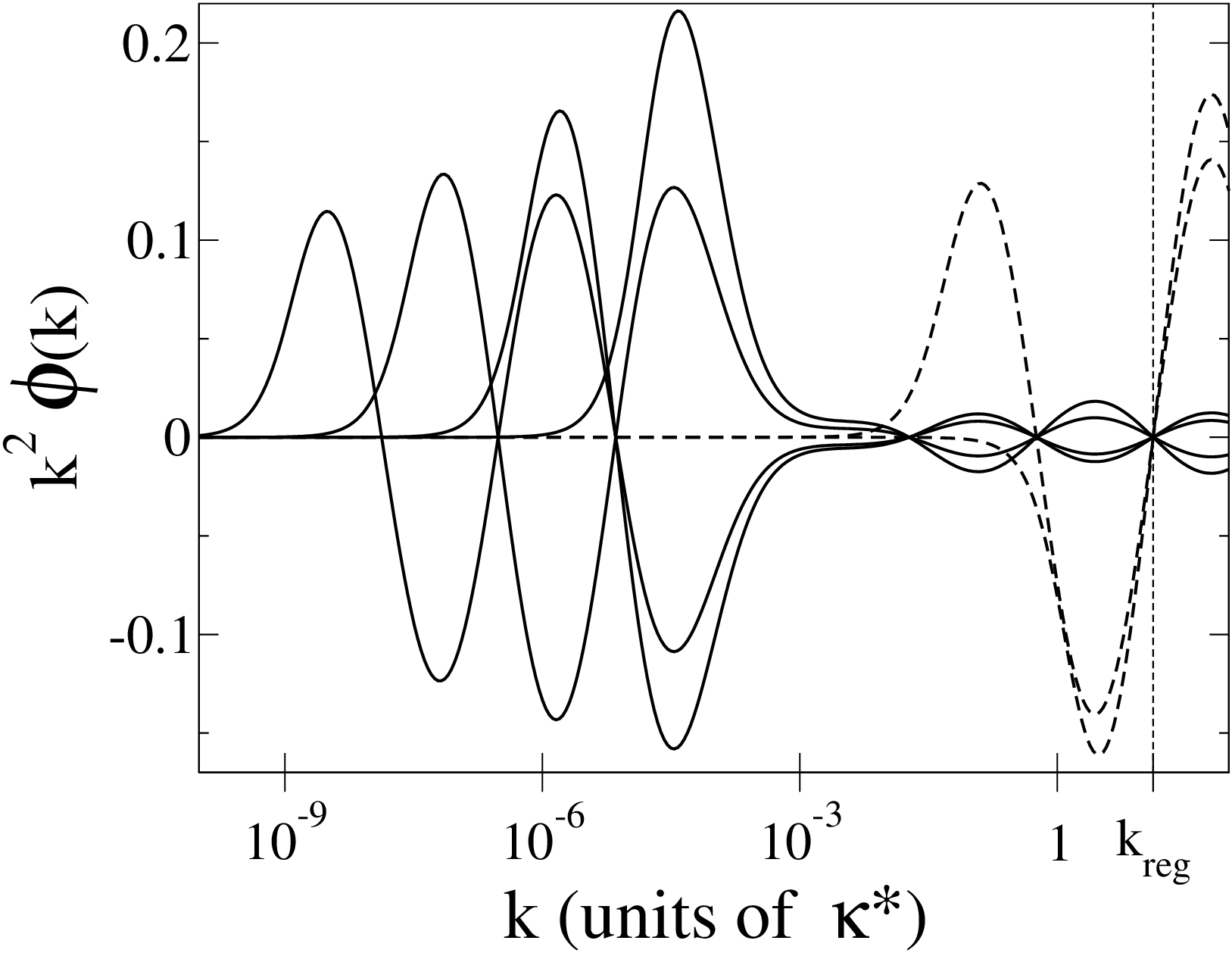}
}
\caption{Eigenfunctions ${k^2\phi_n(k)}$ for the deepest trimers states obtained after normalization of Eq.~\eqref{eq:sko_regular} for ${a_{\rm bg} \kappa^\star=-10^2}$ and ${R^\star \kappa^\star=10^4}$. Solid lines: eigenstates with a binding wavenumber given approximately by the spectrum of a narrow resonance [${q_p = \kappa^\star\,' \exp(p\pi/s_0)}$]. Dashed lines: eigenstates with a binding wavenumber given approximately by the spectrum of a broad resonance [${q_n = \kappa^\star\exp(n\pi/s_0)}$]. Dotted vertical line: position of the node imposed by Eq.~\eqref{eq:filtre}.}
\label{fig:function_cross}
\end{center}
\end{figure}
For a vanishingly small background scattering length ${(|a_{\rm bg}| \to 0)}$ and a large width radius, the spectrum obtained from Eq.~\eqref{eq:sko_regular} coincides with the first one. As ${|a_{\rm bg}|}$ grows at a fixed and large value of the width radius [Fig.~(\ref{fig:Spec_Rstar_mabg}-a)], deeper 'broad-type' trimers (${q \gg \kappa^\star\,'}$) appear in the spectrum and 'narrow-type' states evolve gradually towards the 'broad type' states.  Interestingly by noticing that the high momentum behavior of the scattering amplitude is given by:
\begin{equation}
\frac{1}{f(k)} \operatornamewithlimits{=}_{k\to \infty} - \frac{1}{a_{\rm bg}} -ik + O(k^{-2}) ,
\label{eq:highfk}
\end{equation}
one can deduce from the universal theory the threshold of apparition of the 'deepest' trimer  by using the formal substitution ${(a \leftrightarrow a_{\rm bg})}$: ${\kappa^\star a_{\rm bg}=-1.51}$ (value of ${\kappa^\star a_*'}$ in Ref.~\cite{Bra06,Gog08,Pri10a}). Second and third thresholds for the apparition of the two others 'deep' trimers in Fig.~(\ref{fig:Spec_Rstar_mabg}-a) are deduced from the first threshold value by the discrete symmetry of scaling factor ${e^{{n\pi}/{s_0}}}$. As shown in Figs.~(\ref{fig:function_cross}-\ref{fig:function_cross2_log}) 'broad type' states are located in a momentum region ${k \gtrsim 1/|a_{\rm bg}|}$ (this is a consequence of Eq.~\eqref{eq:highfk}) and  'narrow type' states  are located in the region ${k \lesssim \kappa^\star\,'}$. Therefore, they experience a very small coupling and this explains the tiny avoided crossings in Fig.~(\ref{fig:Spec_Rstar_mabg}-a).

\begin{figure}[hx] 
\begin{center}
\resizebox{8 cm}{!}
{
\includegraphics{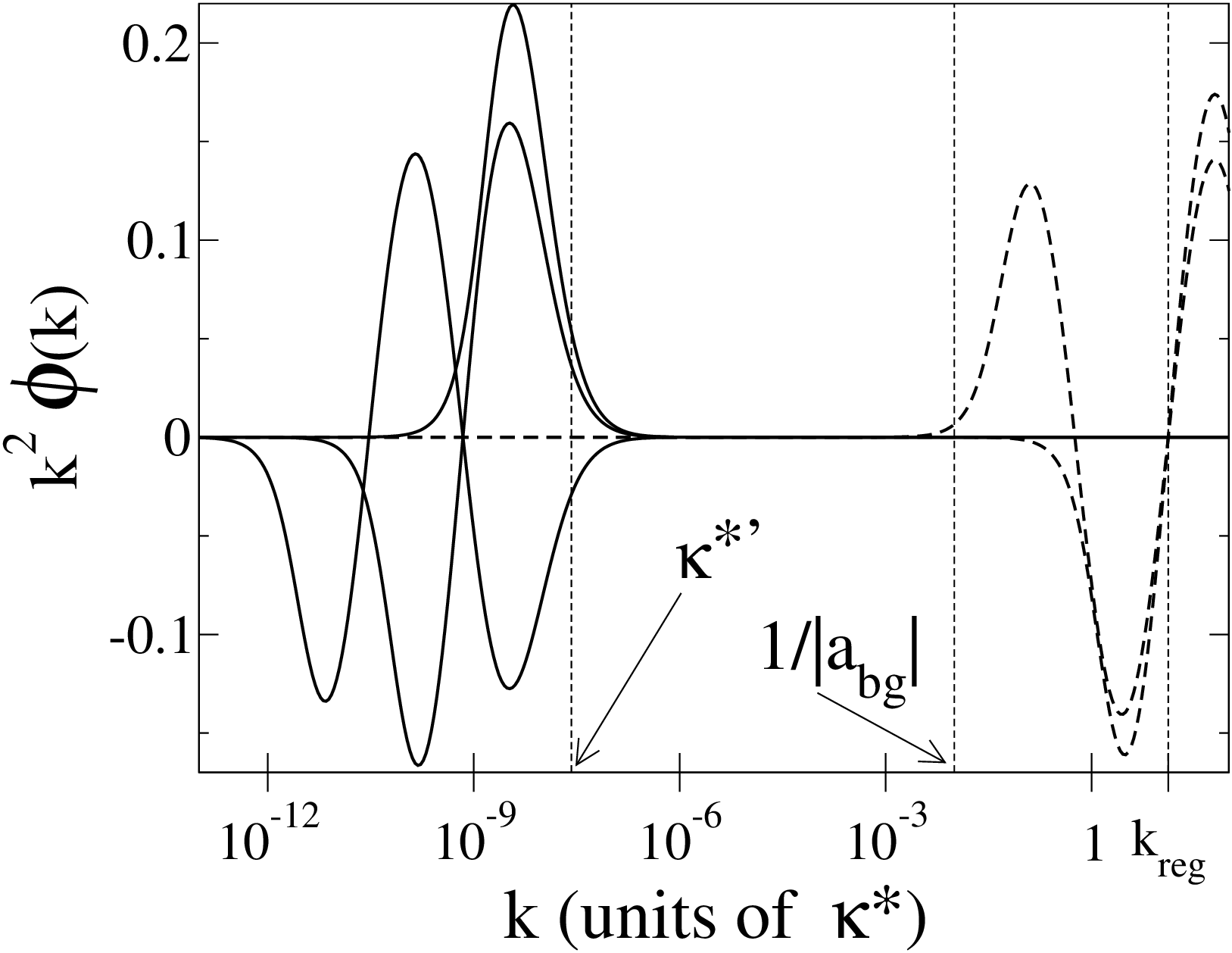}
}
\caption{Same as Fig.~(\ref{fig:function_cross}), but for ${a_{\rm bg} \kappa^\star=-10^2}$ and ${R^\star \kappa^\star=10^8}$. In this regime, characterized by the inequality ${R^\star \gg |a_{\rm bg}|}$, there is a clear separation of scale for the two type of eigenstates: the deepest one of the 'broad type' are located in the momentum region ${k\gtrsim 1/|a_{\rm bg}|}$, those of the 'narrow type' are located in the momentum region ${k \lesssim \kappa^\star\,'}$.}
\label{fig:function_cross2}
\end{center}
\end{figure}

\begin{figure}[hx] 
\begin{center}
\resizebox{8 cm}{!}
{
\includegraphics{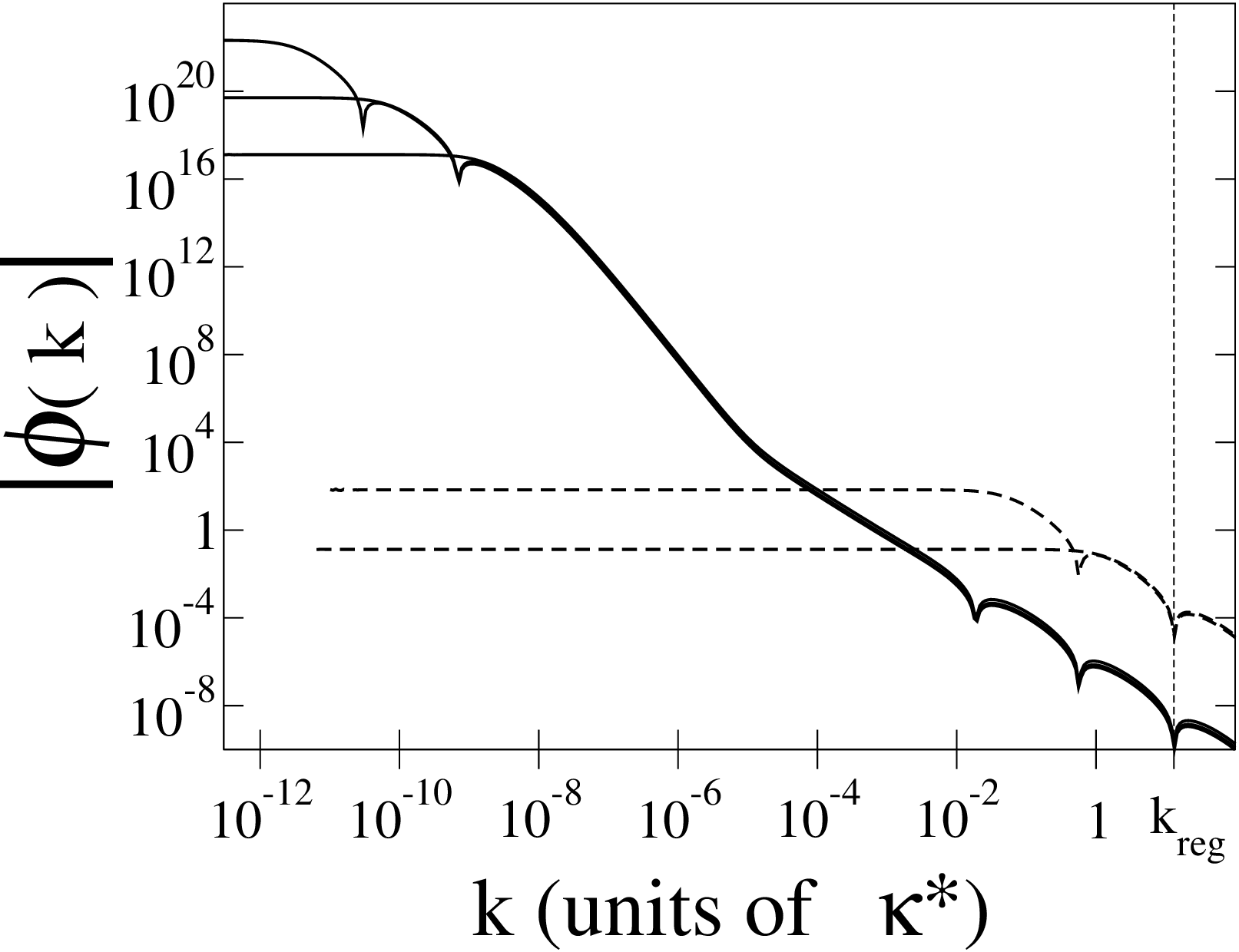}
}
\caption{Absolute value of the eigenfunctions $|\phi(k)|$ in Fig.~(\ref{fig:function_cross2}) (${a_{\rm bg} \kappa^\star=-10^2, \,R^\star \kappa^\star=10^8}$) in logarithmic scale. The figure, shows that even in this regime where 'broad type' and 'narrow type' eigenfunctions are well separated, eigenfunctions share the same 0 at $k=k_{\rm reg}$. 'Narrow type' eigenstates experience a ${k^{-4}}$ power law in the region $1 \ll k R^\star \ll \sqrt{\frac{R^\star}{|a_{\rm bg}|}}$.}
\label{fig:function_cross2_log}
\end{center}
\end{figure}

\subsection{Fano-Efimov resonances}

In the case of a positive background scattering length ${a_{\rm bg}>0}$, the model supports a dimer with a binding wave number ${q_{\rm dim}=(1+\sqrt{1+4 a_{\rm bg}/R^\star})/2a_{\rm bg}}$. For ${q>q_{\rm dim}}$, one finds eigenstates of the 'broad type'. For ${q<q_{\rm dim}}$ there is an atom-dimer continuum, and in what follows Eq.~(\ref{eq:Skorniakov}) is solved for a $s$-wave atom-dimer scattering process. For an incoming atom-dimer plane wave of momentum $\mathbf k_0$, the wavenumber $q$ is given by ${q=\sqrt{q_{\rm dim}^2-3k_0^2/4}}$ and the $s$-wave scattering ansatz is: 
\begin{equation}
\phi(\mathbf k)= 2\pi^2 \frac{\delta(k - k_0)}{k_0^2} + \frac{4\pi g(k)}{k^2-k_0^2-i0^+}.
\label{eq:ansatz}
\end{equation}
One recognizes in Eq.~\eqref{eq:ansatz} the $s$-wave atom-dimer scattering amplitude defined by ${f_{\rm ad}(k_0)=g(k_0)}$. Injecting Eq.~\eqref{eq:ansatz} in Eq.~\eqref{eq:Skorniakov}, one obtains:
\begin{multline}
\frac{{\mathcal K}_q(k,k_0)}{2 k_0^2} [1+ik_0 g(k_0)]+ {\cal P} \int_0^\infty\!\! \frac{du}{\pi} \, 
\frac{{\mathcal K}_q(k,u)
}{u^2-k_0^2} g(u)=\\
\frac{3 g(k)}{8(q_{\rm dim}+q_{\rm col})} \left[1+\frac{(1+\frac{q_{\rm col}}{q_{\rm dim}})(1 -\frac{1}{a q_{\rm dim}})}{R^\star a_{\rm bg}(1-\frac{a_{\rm bg}}{a}) q_{\rm col}^2-1} \right] .
\label{eq:diff}
\end{multline}
\begin{figure}[h] 
\resizebox{8 cm}{!}
{
\includegraphics{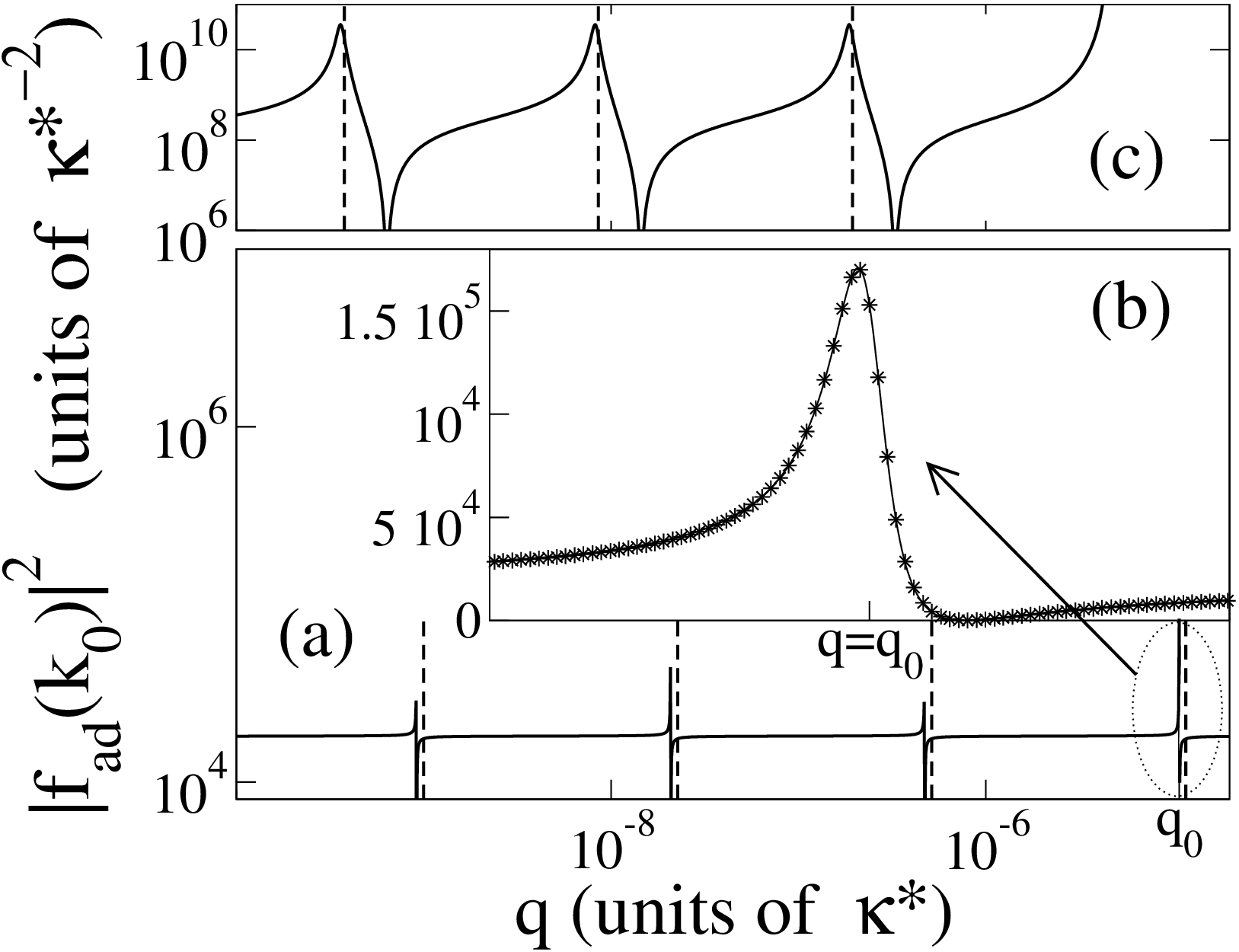}
}
\caption{Square modulus of the $s$-wave atom-dimer scattering amplitude in the resonant regime (${|a|=\infty, a_{\rm bg}>0}$) as a function of the wave number ${q=\sqrt{q_{\rm dim}^2-3k_0^2/4}}$, where ${k_0}$ is the collisional momentum and ${q_{\rm dim}}$ is the dimer's binding wave number. The plot is in units of ${\kappa^\star}$. (a): ${R^\star\kappa^\star=10^4}$ and ${a_{\rm bg}\kappa^\star=5\times10^2}$; dashed vertical lines: 'narrow-state' spectrum from Refs.~\cite{Pet04b,Gog08}. (b): Details of the resonant peak at ${q=q_0}$ in Fig.~(a); stars: results of the computation; solid line: fit using the Fano profile. (c): ${R^\star\kappa^\star=10^4}$ and ${a_{\rm bg}\kappa^\star=5\times10^6}$; dashed vertical lines: spectrum of the universal theory~\cite{Efi70,Bra06}.}
\label{fig:quasi_efi}
\end{figure}
In Eq.~\eqref{eq:diff}, the integral is performed in the sense of Cauchy principal value and the filtered equation is obtained by making the substitution ${{\mathcal K}_q(k,v) \to {\mathcal K}_q(k,v)-{\mathcal K}_q(k_{\rm reg},v)}$. As shown in Fig.~(\ref{fig:quasi_efi}), the atom-dimer scattering amplitude obtained after regularization of Eq.~\eqref{eq:diff} supports a series of peaks as a function of the wavenumber $q$ at momenta ${q_{\rm trim}}$ related by the Efimov scaling factor ${e^{\pi/s_0}}$. In the regime where the peaks are well separated, their profile are typical of Fano resonances wich can be parameterized by the following expression \cite{Fan61}:
\begin{equation}
|f(k_0)|^2=|f_0|^2 \frac{(q_F+\epsilon)^2}{1+\epsilon^2} ,
\label{eq:Fano}
\end{equation}
where ${q_F}$ is the 'profile parameter' which is related to the asymmetry of the peak, ${\epsilon={2(q^2-q^2_{\rm trim})}/{\gamma}}$ is the reduced energy, $q_{\rm trim}$ is the binding wave number of the excited Efimov state, and ${\gamma}$ is the width of the resonance. Interestingly, for a fixed value of ${(a_{\rm bg}, R^\star)}$ the peaks computed in Fig.~(\ref{fig:quasi_efi}) are characterized approximately by the same ratio $\gamma/q^2_{\rm  trim}$. As expected, in the limit where ${R^\star \gg a_{\rm bg} \gg 1/\kappa^\star}$ excited trimers are decoupled from the high momentum scale $\kappa^\star$. First, the peaks are located  at the positions predicted by the effective range approach and second, their width goes to 0 as $R^\star$ increases. This can be interpreted from the existence of a large scale separation between quasi-bound trimers extending over the region ${k \lesssim 1/R^\star}$ and the dimer of momentum ${q_{\rm dim} \gg 1/ R^\star}$. Hence the production of long-lived Efimov's trimers seems to be possible in this regime.  If one takes into account only spontaneous decay toward one atom and one dimer, the trimer's lifetime is given by: ${\tau_D=m/(\gamma \hbar)}$. As an example,  the peak in Fig.~(\ref{fig:quasi_efi}-b) at ${q_{\rm trim}=q_0}$ was computed for ${R^\star \kappa^\star=10^4\gg a_{\rm bg}\kappa^\star =5\times10^2\gg1}$, and one finds ${\gamma \simeq 2\times10^{-3} q_0^2}$. In this case the lifetime of the quasi-bound state is limited by inelastic processes with other components in the system. In the opposite limit where $a_{\rm bg}\gg R^\star$ [see Fig.~(\ref{fig:quasi_efi}-c)], resonances are large meaning that the three-body quasi-bound state are not long-lived and peaks are located at the positions predicted by the universal theory. For negative background scattering length ${a_{\rm bg}<0}$, the decay rate of 'narrow type' trimers involves deep dimer state which are not described in the present model. In the regime where ${R^\star \gg |a_{\rm bg}| \gg 1/\kappa^\star}$, the eigenfunction $\phi(k)$ experience a $1/k^4$ decreasing law in the region ${1 \ll kR^\star \ll \sqrt{R^\star/|a_{\rm bg}|}}$ where ${1/f(ik) \sim R^\star k^2}$ [see  Fig.~\eqref{fig:function_cross2_log}]. This suggests that the coupling with deep dimers (not supported by the model) is small. One thus expects also large lifetimes in this limit. A possible way to observe long-lived trimers is given by the radio-frequency association technique which has been already implemented in fermionic systems \cite{Lom10}.

For finite background scattering length, the present zero-range model is justified in the limit where ${|a_{\rm bg}|\gg b}$. When this last condition is not satisfied, more realistic two-channel model as in Refs.~\cite{Lee07,Jon10} should be used in order to evaluate the lifetime of the trimers. Work in this direction is in progress. 

\section*{CONCLUSION}

In this paper, a zero range model is introduced for solving the three-boson problem in the regime of  resonant $s$-wave binary interactions.  At the two body level, the model takes into account the scattering length ${a}$, the width radius ${R^\star}$ (related to the width of a magnetic Feshbach resonance),  and the background (or off-resonant) scattering length ${a_{\rm bg}}$. Effective range \cite{Pet04b} and Bethe Peierls \cite{Bet35} approaches can thus be obtained as specific limits of this formalism. For three interacting bosons, a generalized Skorniakov Ter-Martirosian equation is derived. The Danilov's filtering condition that fixes the value of the three-body parameter ${\kappa^\star}$ is imposed through a nodal condition.

The trimers' spectrum  is then solved at resonance (${|a|=\infty}$). For negative background scattering length ${a_{\rm bg}<0}$, as one varies $R^\star$ or $a_{\rm bg}$ for a fixed value of the three-body parameter (${\kappa^\star}$), the spectrum exhibits a crossover between Efimov states of narrow resonances (characterized by an effective three-body parameter ${\kappa^\star\,'}$ defined by $R^\star$ only, with ${\kappa^\star\,'\simeq 2.65/R^\star}$) and Efimov states of broad resonances with binding wave numbers deduced from ${\kappa^\star}$ by the scaling factor ${e^{\pi/s_0}}$. For a positive background scattering length ${a_{\rm bg}>0}$, a dimer exists. For energies higher than the dimer's energy, Efimov states are quasi-bound and manifest themselves as Fano resonances. For sufficiently large values of $R^\star$ these Fano resonances are narrow and the peaks are located at the positions predicted by the effective range theory. In the latter regime, quasi-bound Efimov trimers thus appear as long lived molecules. 

More generally, this result opens the issue of the stability of resonant many-boson systems in the regime of narrow resonances.

\section*{ACKNOWLEDGEMENTS} 

Y. Castin, M. Jona-Lasinio, C. Mora and F. Werner are acknowledged for discussions. The Laboratoire de Physique Th\'{e}orique de la Mati\`{e}re Condens\'{e}e is UMR 7600 of the CNRS and its Cold Atoms group is associated with the IFRAF.


\begin{thebibliography}{99}
\bibitem{LowE} A low energy scale refers to energies associated with a momentum $k$ such that ${|k| b \ll 1}$.
\bibitem{Rstar} If $R^\star_{\rm fit}$ is of the order of $1/\kappa^\star$, results  similar to those presented here should be obtained.
\end{thebibliography}
\end{document}